\documentclass[12pt]{iopart}
\usepackage{ulem}
\usepackage{color}
\usepackage{graphicx}
\usepackage{epsfig}
\usepackage{bm}
 \newcommand{\be}{\begin{equation}}
\newcommand{\ee}{\end{equation}}
\newcommand{\bea}{\begin{eqnarray}}
\newcommand{\eea}{\end{eqnarray}}

\begin{document}
\title{Heavy quark diffusion in pre-equilibrium stage of heavy ion collisions} 
\author{Santosh K. Das$^{1,2}$, Marco Ruggieri$^{1}$, Surasree Mazumder$^{3}$, Vincenzo Greco$^{1,2}$ and Jan-e Alam$^{3}$}

\address{$^1$ Department of Physics and Astronomy, University of Catania, Via S. Sofia 64, I-
95125 Catania, Italy}
\address{$^2$ Laboratori Nazionali del Sud, INFN-LNS, Via S. Sofia 62, I-95123 Catania, Italy}
\address{$^3$ Variable Energy Cyclotron Centre, 1/AF, Bidhan Nagar , Kolkata - 700064}
\ead{santosh@lns.infn.it}
\date{\today}
\begin{abstract}
The drag and diffusion coefficients of heavy quarks (HQs) have been evaluated in the 
pre-equilibrium phase of the evolving fireball produced in heavy ion collisions at 
RHIC and LHC energies. The KLN and classical Yang-Mills spectra have been used
for describing the momentum distributions of the gluons produced just after the 
collisions but before they thermalize.  The interaction 
of the HQs with these  gluons has been treated within the 
framework of perturbative QCD. We have observed  that the HQs are dragged almost equally
by the kinetically equilibrated and out-of-equilibrium gluonic systems.  
We have also noticed  that the HQs diffusion  
in the pre-equilibrium gluonic phase is as fast as in the kinetically equilibrated gluons.
Moreover, the diffusion is faster in the pre-equilibrium phase 
than  in the  chemically equilibrated quark-gluon plasma. 
These findings may have significant impact on the analysis of experimental 
results on the elliptic flow and the high momentum suppression of the 
open charm and beauty hadrons. 

\vspace{2mm}
\end{abstract}
  \vspace{2mm}
\maketitle

\section{Introduction}
The  primary intent of the ongoing nuclear collision programmes at 
the Relativistic Heavy Ion Collider (RHIC) and Large Hadron Collider (LHC) energies 
is to create a new state of matter called  Quark Gluon Plasma (QGP).
The bulk properties of this state  is  governed by the light quarks ($q$) and gluons ($g$).  
The heavy quarks  [HQs\,$\equiv$  charm (c)  and beauty (b)] 
are considered  as efficient tools to probe the early 
state of the system~(see \cite{hfr,Rapp:2008tf,Averbeck:2013oga} for a review)
mainly for the following reasons:
(i) HQs are produced at very early stage;
(ii) The probability of creation and annihilation of the  
HQs during the evolution of the fireball is small.
Hence the HQs can witness the entire evolution of
the system as they are created early and
survive the full evolution without being annihilated and/or created.

The data on the suppression of the charm quark at large momentum
($R_{AA}(p_T)$) ~\cite{stare,phenixelat,phenixe,alice}
and their elliptic flow ($v_2$)~\cite{phenixelat,alicev2} have been analyzed 
by several authors (see  {\it e.g.} \cite{hfr,Rapp:2008tf,Averbeck:2013oga} and references therein)
 to characterize the system formed in HIC at RHIC and LHC collisions. 

It  has been argued that the thermalization times of
HQs are larger than the light quarks and gluons~\cite{shuryak,prlja} 
by a factor $\sim M/T$~\cite{moore}, where $M$ is the mass of the HQs and $T$ 
is the temperature of the medium. Therefore, the evolution of
the HQs in the thermal bath of light quarks and gluons can be
treated within the ambit of Brownian motion, although a detailed investigation
of this problem within the framework of Boltzmann equation has revealed that 
the evolution of charm quarks as a Brownian particle requires some corrections~\cite{fs}
(see also ~\cite{Molnar:2006ci,gossiauxv2,gre,you,Uphoff:2014cba,Song:2015sfa}). 
 
Several theoretical attempts have been made 
to study the evolution of the HQs  within the framework of 
Fokker Plank equation
~\cite{hfr,moore,rappv2,hvh,hiranov2,cmko,Das,alberico,jeon,bass,rappprl,ali,hees,qun,lambda,Juan_RAA,Das:2015ana}.  
However, the roles of pre-equilibration phase have been ignored in these works.  An attempt
has been made in the present work to study this role. This is important because the 
HQs are produced in the hard collisions of the colliding partons of the nuclei and
they inevitably interact with the pre-equilibrium phase of the bulk matter.
The effect of the pre-equilibrium phase might be more significant for low-energy nuclear 
collisions: for example, in the case Au+Au collisions at RHIC energy
simulations show that equilibration is achieved approximately within $1$ fm/c,
while the lifetime of the QGP phase turns out to be about $5$ fm/c \cite{Ruggieri:2013bda,Ruggieri:2013ova};
hence the lifetime of the out-of-equilibrium phase is approximately $20\%$ of the
total lifetime of the QGP. 
The lifetime of the QGP phase is
of about 10 fm/c and the duration of the pre-equilibrium phase  is 
about 0.4 fm/c for LHC collision condition. Although, this may suggest the dwindling role of the pre-equilibrium phase 
for collisions at higher energies, estimates of drag and diffusion coefficients of HQs  
done in  this work indicate non-negligible contributions 
of the pre-equilibrium phase.
The motivation of this work is to estimate  the drag and diffusion coefficients of the HQs
interacting elastically with the non-equilibrated gluons constituting
the medium.  

%We evaluate the impact of kinetic non-equilibrium, 
%i.e. non thermal $p_T$ spectra, and chemical non equilibrium  comparing a plasma with only gluons 
%in contrast to one where gluons and quarks follow the equilibrium abundances.
 
The paper is organized as follows. In the next section we discuss the formalism
used to evaluate the drag and diffusion coefficients of the heavy quarks in the
preequilibrium stage.  In Section III we summarize the 
out-of-equilibrium initial conditions we implement in the calculations.
Section IV is devoted to present the results and
section V contains summary and discussions.

\section{Formalism}
The Boltzmann Transport Equation (BTE) describing the evolution of the HQs in the pre-equilibrated gluonic system can be written as:
\bea
\left[\frac{\partial}{\partial t}+\frac{\vec{p}}{E}\cdot\frac{\partial}{\partial \vec{x}}
+\vec{F}\cdot\frac{\partial}{{\partial \vec{p}}}
\right] f(\vec{x},\vec{p},t)=\left[\frac{\partial f}{\partial t}\right]_{collisions}~.
\label{boltzmann}
\eea
%%%%%%%%%%%%%%%%%%%%%%%%%%%%%%%%%%%%%%%%%%%%%%%%%%%%%%%%%%%%%%%%%%%%%%%%%
For binary interaction the collisional integral appearing in the right hand
side of BTE can be written as:
%%%%%%%%%%%%%%%%%%%%%%%%%%%%%%%%%%%%%%%%%%%%%%%%%%%%%%%%%%%%%%%%%%%%%%%%%
\be
\left[\frac{\partial f}{\partial t}\right]_{collisions}= \int d^{3}\vec{k}[w(\vec{p}+\vec{k},\vec{k})
f(\vec{p}+\vec{k})-w(\vec{p},\vec{k})f(\vec{p})].
\label{BTE}
\ee
%%%%%%%%%%%%%%%%%%%%%%%%%%%%%%%%%%%%%%%%%%%%%%%%%%%%%%%%%%%%%%%%%%%%%%%%%%
where $w(\vec{p},\vec{k})$ is the rate of collision which encodes the change of HQs momentum 
from $\vec{p}$ to $\vec{p}-\vec{k}$ is given by,
\be
\omega(p,k)=g_G\int\frac{d^3q}{(2\pi)^3}f^\prime(q)v\sigma_{p,q\rightarrow p-k,q+k}
\label{eq:omega1}
\ee
where $f^\prime$ is the phase space distribution of the particles in the bulk,
$v$ is the relative velocity between the two collision partners,
$\sigma$ denotes the cross section and $g_G$ is the statistical
degeneracy of the particles in the bulk.

Considering only the soft scattering approximation~\cite{BS}, the integro-differential 
Eq.~\ref{BTE} can be reduced to the Fokker Planck equation:
%%%%%%%%%%%%%%%%%%%%%%%%%%%%%%%%%%%%%%%%%%%%%%%%%%%%%%%%%%%%%%%%%%%%%%%%%
\be
\frac{\partial f}{\partial t}= \frac{\partial}{\partial p_{i}}\left[A_{i}(\vec{p})f+\frac{\partial}{\partial p_{j}}[B_{ij}
(\vec{p})f]\right]~~,
\label{landaukeq}
\ee
where the kernels are defined as
\be
A_{i}= \int d^{3}\vec{k}w(\vec{p},\vec{k})k_{i}~~,
\label{eqdrag}
\ee
and
\be
B_{ij}= \frac{1}{2} \int d^{3}\vec{k}w(\vec{p},\vec{k})k_{i}k_{j}~.
\label{eqdiff}
\ee
In the limit $\mid\bf{p}\mid\rightarrow 0$,  $A_i\rightarrow \gamma p_i$
and $B_{ij}\rightarrow D\delta_{ij}$ where $\gamma$ and $D$ 
are the drag and diffusion coefficients 
respectively.
We notice that under the assumption of soft collisions, the non-linear integro-differential equation, Eq.~\ref{boltzmann}
reduces to a much simpler linear partial differential equation, Eq.~\ref{landaukeq},   
provided the  function $f^\prime (q)$ is known.

Now we evaluate the  $\gamma$ and $D$ for HQs interacting  elastically
with the bulk particles in the pre-equilibrium phase that appear in HIC before thermalization.
The $\gamma$ for the process $HQ(p_1) + g(p_2) \rightarrow HQ(p_3) + g(p_4)$ ($p_i$'s are
the respective momenta of the colliding particles)
can be expressed in terms of $A_i$ ~\cite{BS} (see also ~\cite{DKS,vc,Berrehrah:2014tva}) as:
\begin{equation}
\gamma=p_iA_i/p^2~,
\end{equation}
where $A_i$ is given by
\bea
A_i&=&\frac{1}{2E_{p_1}} \int \frac{d^3p_2}{(2\pi)^3E_{p_2}} \int \frac{d^3p_3}{(2\pi)^3E_{p_3}}
\int \frac{d^3p_4}{(2\pi)^3E_{p_4}}  \nonumber \\ 
&&\frac{1}{g_{HQ}} 
\sum  {\overline {|{\cal{M}}|^2}} (2\pi)^4 \delta^4(p_1+p_2-p_3-p_4) 
\nonumber \\
&&{f}(p_2)\{1\pm f(p_4)\}[(p_1-p_3)_i] \equiv \langle \langle  
(p_1-p_3)_i\rangle \rangle . \nonumber \\ 
\label{eq1}
\eea
where  $g_{HQ}$ denotes the statistical degeneracy of HQ, 
$f(p_2)$ is  the momentum distributions of the incident particles,
$1 \pm f(p_4)$ is  the 
momentum distribution  (with Bose enhancement or Pauli suppression) 
of the final state particles in the bath and
${\overline {|{\cal{M}}|^2}}$ represents the square modulus of the spin averaged 
invariant amplitude for the elastic process, evaluated here using  pQCD.  
The drag coefficient in Eq.~(\ref{eq1}) is the measure of the average of the momentum transfer, $p_1-p_3$, 
weighted  by the interaction through  $\overline{|{\cal{M}}|^2}$.
Similarly the momentum diffusion coefficient, $D$, can be defined as:
\begin{equation}
D=\frac{1}{4}\left[\langle \langle p_3^2 \rangle \rangle -
\frac{\langle \langle (p_1\cdot p_3)^2 \rangle \rangle }{p_1^2}\right]
\label{eq3}~,
\end{equation}
and it represents the averaged square momentum transfer (variance) through the interaction process mentioned above.

The following  general expression has been used to 
evaluate the drag  and diffusion coefficients numerically with an appropriate choice of $Z(p)$, 
\bea
&&\ll Z(p)\gg=\frac{1}{512\pi^4} \frac{1}{E_{p_1}} \int_{0}^{\infty} 
\frac{p_2^2 dp_2 d(cos\chi)}{E_{p_2}} 
\nonumber \\
&&~~~\hat{f}(p_2)\{ 1\pm f(p_4)\}\frac{\lambda^{\frac{1}{2}}(s,m_{p_1}^2,m_{p_2}^2)}{\sqrt{s}} 
\int_{1}^{-1} d(cos\theta_{c.m.})  
\nonumber \\ 
&&~~~~~~~~~~~~~~~~~~
\frac{1}{g_{HQ}} \sum  {\overline {|{\cal{M}}|^2}} \int_{0}^{2\pi} d\phi_{c.m.} Z(p) \nonumber \\ 
\label{transport}
\eea
where $\lambda(s,m_{p_1}^2,m_{p_2}^2)=\{s-(m_{p_1}+m_{p_2})^2\}\{s-(m_{p_1}-m_{p_2})^2\}$. 

\section{Initial conditions}
In most of the earlier works the distribution functions
in Eq.~(\ref{transport}) are taken as
equilibrium distribution, {\it i.e.} Fermi-Dirac for  quarks and
anti-quarks and Bose-Einstein for gluons. 
The transport coefficients are then corresponding to
the motion of the HQs in a thermalised medium assumed to be formed
within the time scale $\sim 1$ fm/c. 

In this article instead, we will evaluate the drag and diffusion coefficients of HQ propagating through a 
non-thermal gluonic system.
We achieve this by substituting the distribution functions 
in Eq.~(\ref{transport}) by  pre-equilibrium distribution of gluons to be specified later.
We choose the normalization of the non-equilibrium distributions
in such a  way that the gluon density for the chosen distribution
and the thermal distribution at initial temperature coincides. 
The initial temperature, $T_i$ is taken as 
$0.34$ GeV for RHIC and $0.51$ GeV for LHC collision conditions.
These initial temperatures are chosen from simulations~\cite{Ruggieri:2013bda,Ruggieri:2013ova,Niemi:2011ix}
done to reproduce $v_2$ and spectra of light hadrons, as well as the $R_{AA}$ and
$v_2$ of heavy mesons~\cite{fs,Das:2015ana}. 
We compare the drag and diffusion coefficients of HQs in the pre-equilibrium era with those
in the thermalized era at the initial temperatures just mentioned.

We now specify the out-of-equilibrium gluon distribution used in this work for evaluating HQs drag and diffusion
coefficients.
According to the general understanding of the dynamics of pre-equilibrium stage
the initial strong gluon fields (the glasma) shatters into gluon quanta in a time scale
which is of the order of the inverse of the saturation scale $Q_s$;
therefore any model of the pre-equilibrium stage will include a $Q_s$.
The first one we consider
is the classical Yang-Mills (CYM) gluon spectrum (see~\cite{Schenke:2013dpa} for details), 
which assumes the initial gluon fields can be expanded in terms
of massless gluonic excitations. 

Beside CYM we also  consider the model known as factorized KLN 
model~\cite{Drescher:2006ca,Hirano:2009ah} which includes the saturation scale
in an effective way through the unintegrated gluon distribution functions which,
for the nucleus $A$, that participates in $A+B$ collision reads:
\begin{equation}
\phi_A\left(x_A,\bm k_T^2,\bm x_\perp\right) =
\frac{1}{\alpha_s(Q_s^2)}\frac{Q_s^2}{\it{max}(Q_s^2,\bm k_T^2)}~
\end{equation} 
a similar equation holds for nucleus $B$. 
The momentum space gluon distribution is then given by
\begin{eqnarray}
\frac{dN}{dy d\bm p_T}&=&
\frac{{\cal N}}{p_T^2}\int d^2 x_T\int_0^{p_T}d\bm k_T
\alpha_s(Q^2)\nonumber\\
&&\times\phi_A\left(x_A,\frac{(\bm k_T + \bm p_T)^2}{4},\bm x_\perp\right)\nonumber\\
&&\times\phi_B\left(x_B,\frac{(\bm k_T - \bm p_T)^2}{4},\bm x_\perp\right)~,
\label{eq:3}
\end{eqnarray}
where ${\cal N}$ is an overall constant which is fixed by the multiplicity.

%%%%%%%%%%%%%%%%%%%%%%%%%%%%%%%%%%%%%%%%%%%%%%%%%%%%%%%
\begin{figure}[t!]
\begin{center}
\includegraphics[width=17pc,clip=true]{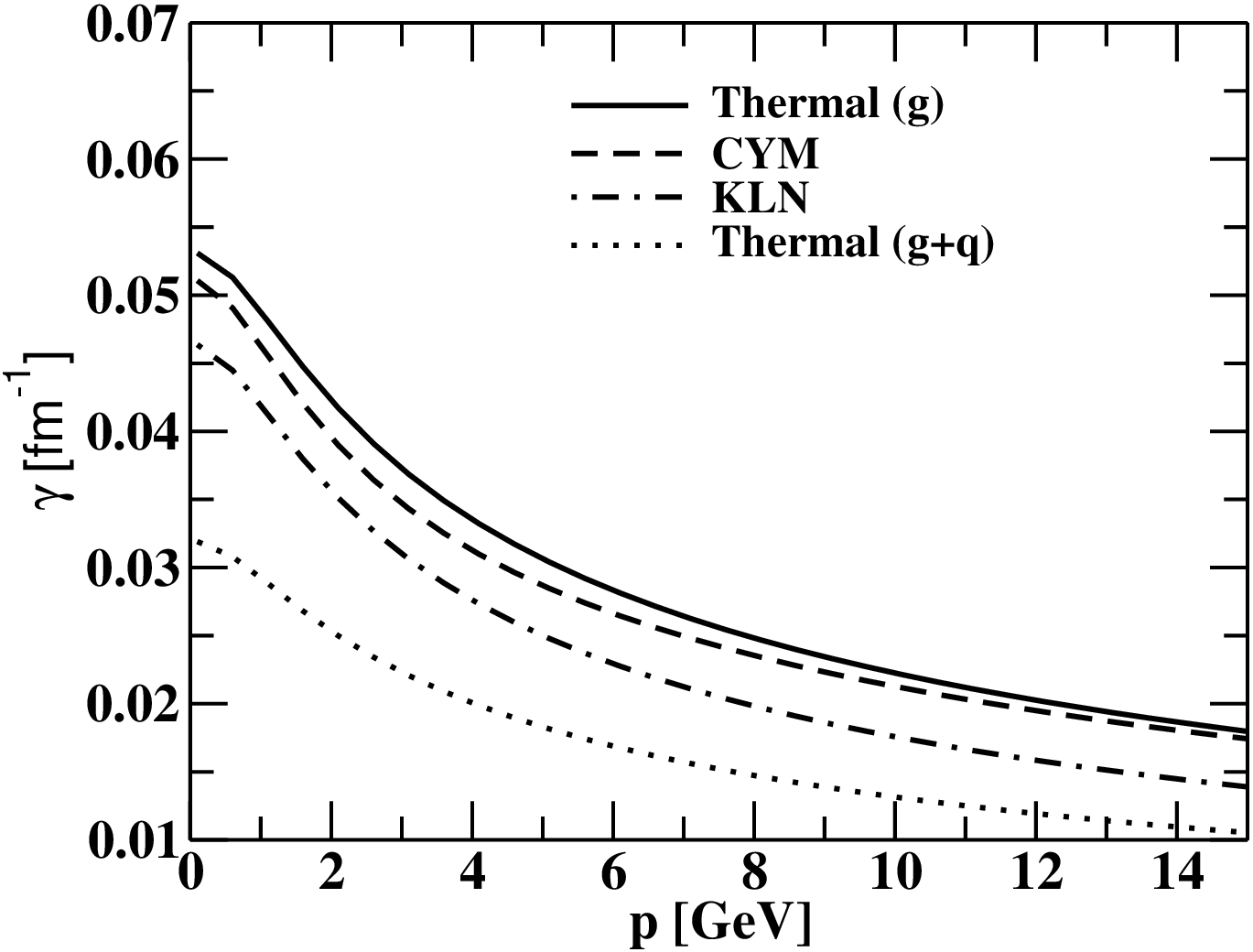}\hspace{2pc}~
\includegraphics[width=17pc,clip=true]{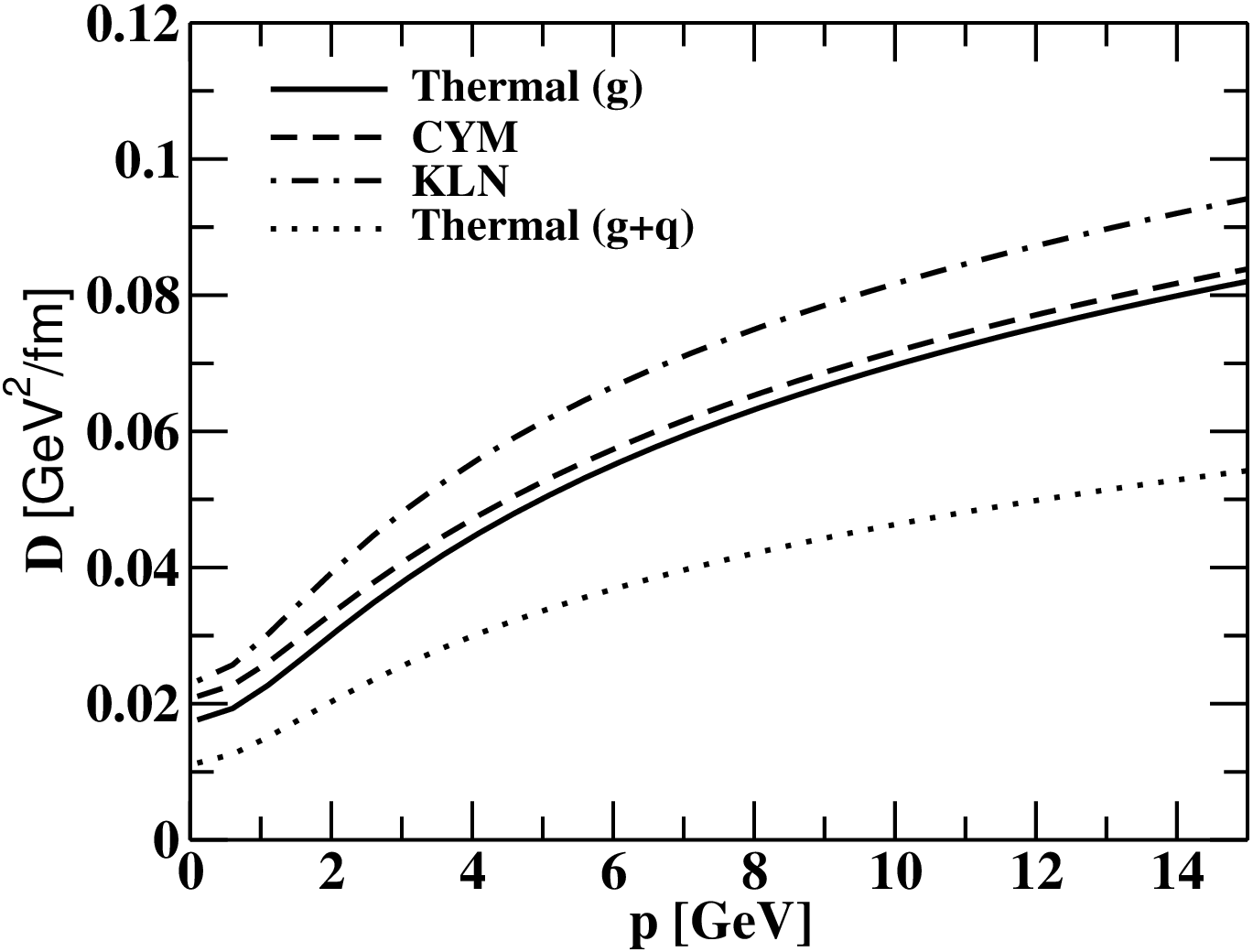}\hspace{2pc}
%\begin{minipage}[b]{14pc}described
\caption{Drag coefficient (left panel) and diffusion coefficient (right panel)
as a function of momentum for charm quark at RHIC energy. 
The results corresponding to thermal gluons (kinetic equilibrium) and thermal quarks and 
gluons (chemical equilibrium) are evaluated at a temperature 340 MeV.
}
%\end{minipage}
\label{fig1}
\end{center}
\end{figure}
%%%%%%%%%%%%%%%%%%%%%%%%%%%%%%%%%%%%%%%%%%%%%%%%%%%%%%%

\section{Results and discussions}
We have evaluated 
the drag and diffusion coefficients of HQs propagating through a system of out-of-equilibrium 
gluons formed at the  very early era of HIC at the RHIC and LHC 
energies. The magnitudes of the  transport coefficients evaluated in the pre-equilibrium  era
 are compared  with those obtained in  the equilibrium phase 
(both kinetic and chemical)  keeping the number of particles fixed
in both  the cases as mentioned above. 
The temperature dependence given in Ref.~\cite{zantow}  has been used to estimate the value
of the strong coupling,  $\alpha_s$ at $T=T_i$.
The Debye screening mass, $m_D=\sqrt{8\alpha_s(N_c+N_f)T^2/\pi}$, 
for a system at temperature, $T$ with $N_c$ colors and $N_f$ flavors
has been used to shield the infra-red divergence associated with the $t$-channel 
scattering amplitude.   $m_D$ is estimated at $T=0.34$ GeV for RHIC energy and 
at $T=0.51$ GeV for LHC energy.  
For the pre-equilibrium and equilibrium system same values of $m_D$  and 
$\alpha_s$ have been used. 
Later in this section we will show results where $m_D$  is estimated
in a self-consistent way with the underlying distribution function.

In the left panel of Fig.~\ref{fig1} we plot the drag coefficient 
of the charm quark as a function of momentum in the pre-equilibrium phase for CYM and KLN
gluon distributions and compared the results with the equilibrated phase (both kinetic and chemical) 
at T=0.34 GeV. 
We find that the magnitude of the drag coefficient in the pre-equilibrium 
phase is quite large, indeed comparable to the one in the kinetic equilibrium phase, 
indicating 
substantial amount of interaction of the charm quarks with the pre-equilibrated gluons. 
We notice that $\gamma$ for the case of chemically equilibrated QGP (dotted line) 
is smaller than the one of the purely gluonic system (solid line in Fig.~\ref{fig1}).
This can be understood  by considering the fact that 
to keep the total number of bath particles fixed some of the gluons in the purely gluonic system has
to be replaced by quarks in the chemically equilibrated QGP. 
Because the gluon appears in more colours than quarks the cross section for $cg$ interaction 
is larger than $cq$ which leads to larger drag of HQs in a pre-equilibrated gluonic system than 
in a chemically equilibrated QGP. 
The CYM distribution gives larger value of the drag coefficient 
compared to KLN distribution. The KLN provides harder momentum distribution (compared to CYM)
hence have a smaller difference with the HQs distribution.  Therefore, the 
momentum transfer between the gluon (with KLN distribution) and the HQ is small.  As the 
drag coefficient is a measure of the momentum transfer weighted by the interaction strength,
the KLN gives rise to lower drag compared to the  one obtained with the CYM distribution.  

The variation of $D$ with momentum for charm quarks has been depicted in 
the right panel of Fig.~\ref{fig1}. 
We find that the magnitude of the $D$ for CYM initial condition is similar 
to the case where the gluons are kinetically  equilibrated.
However, the $D$ for the KLN initial  condition is larger than CYM. 
As mentioned before, the KLN has harder momentum distributions than the
CYM, hence KLN distribution 
has larger variance compared to CYM. Since the momentum diffusion 
is a measure of the variance acquired through interaction with
the bulk,  the charm diffusion coefficient is larger for KLN distribution,
which is reflected in the results displayed in Fig.~\ref{fig1}.

%%%%%%%%%%%%%%%%%%%%%%%%%%%%%%%%%%%%%%%%%%%%%%%%%%%%%%%
\begin{figure}[t!]
\begin{center}
\includegraphics[width=17pc,clip=true]{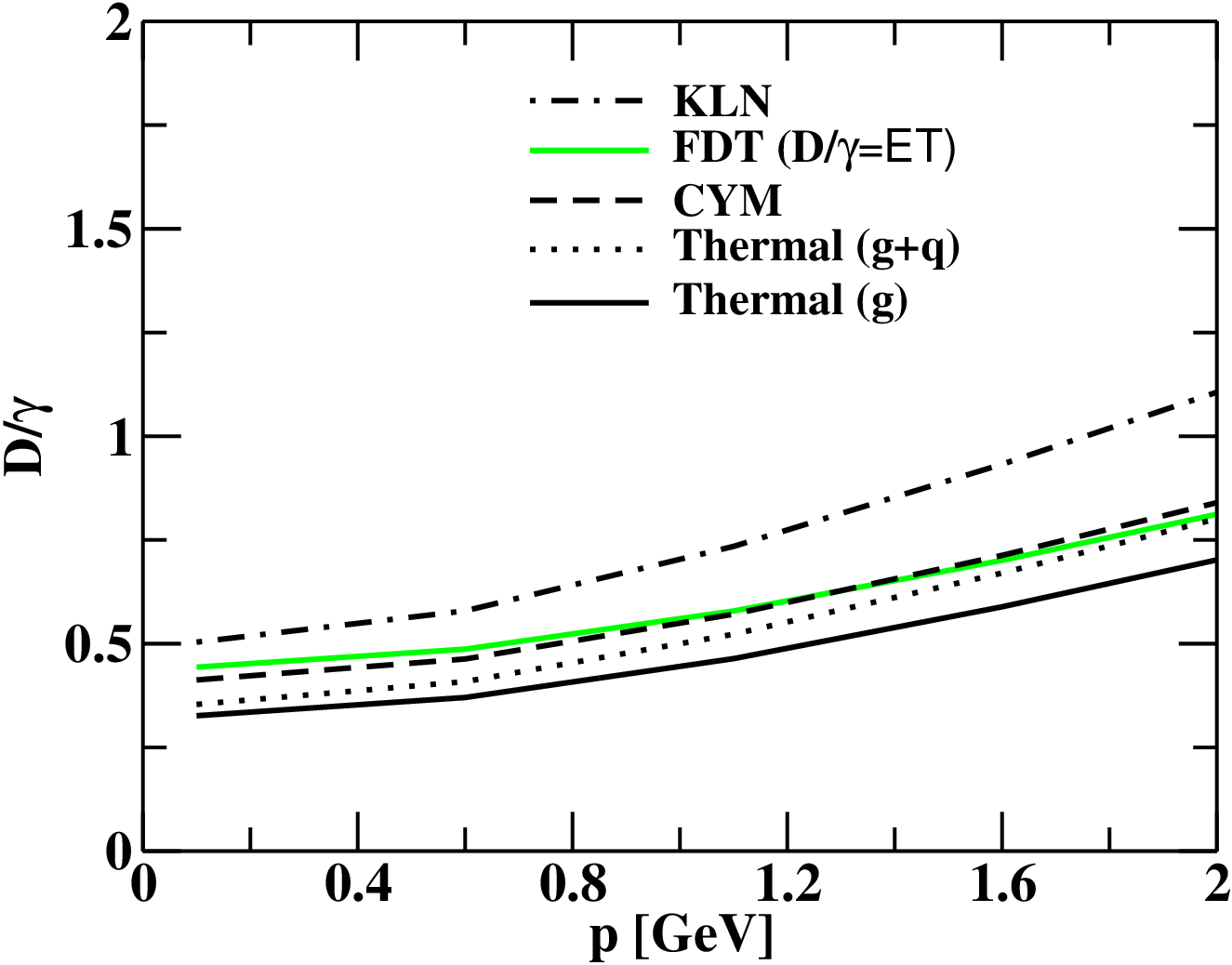}\hspace{2pc}
%\begin{minipage}[b]{14pc}described
\caption{$D/\gamma$  as a function of momentum for charm quark at RHIC energy. 
The result corresponding to equilibrium cases  are evaluated at a temperature 340 MeV.
}
%\end{minipage}
\label{fig300}
\end{center}
\end{figure}
%%%%%%%%%%%%%%%%%%%%%%%%%%%%%%%%%%%%%%%%%%%%%%%%%%%%%%%

In Fig.\ref{fig300}, we have depicted the diffusion to drag ratio, $D/\gamma$ as function of momentum.
$D/\gamma$ can be used to understand the deviation of the calculated values from the value
obtained by using Fluctuation-Dissipation theorem (FDT) 
(green line in the figure). 
Since the KLN has a harder momentum distribution, results obtained from
KLN input deviates from  FDT more.

In this work, the dependence of the drag/diffusion coefficient on the variation of the 
phase space distribution is addressed. To make the present study 
more consistent the effects of phase space on the dynamics through 
Debye screening mass has been included. It is well-known that the  Debye mass depends on the
equilibrium distribution as follows:
\be
m_D^2=\pi\alpha_s g_G \int \frac{d^3p}{(2\pi)^2} \frac{1}{p} (N_cf_g+N_ff_q)
\label{mdd}
\ee
It is interesting
to see how the results are affected when the equilibrium distribution 
in Eq.~\ref{mdd} is replaced by the KLN or CYM distributions.

%%%%%%%%%%%%%%%%%%%%%%%%%%%%%%%%%%%%%%%%%%%%%%%%%%%%%%%
\begin{figure}[t!]
\begin{center}
\includegraphics[width=17pc,clip=true]{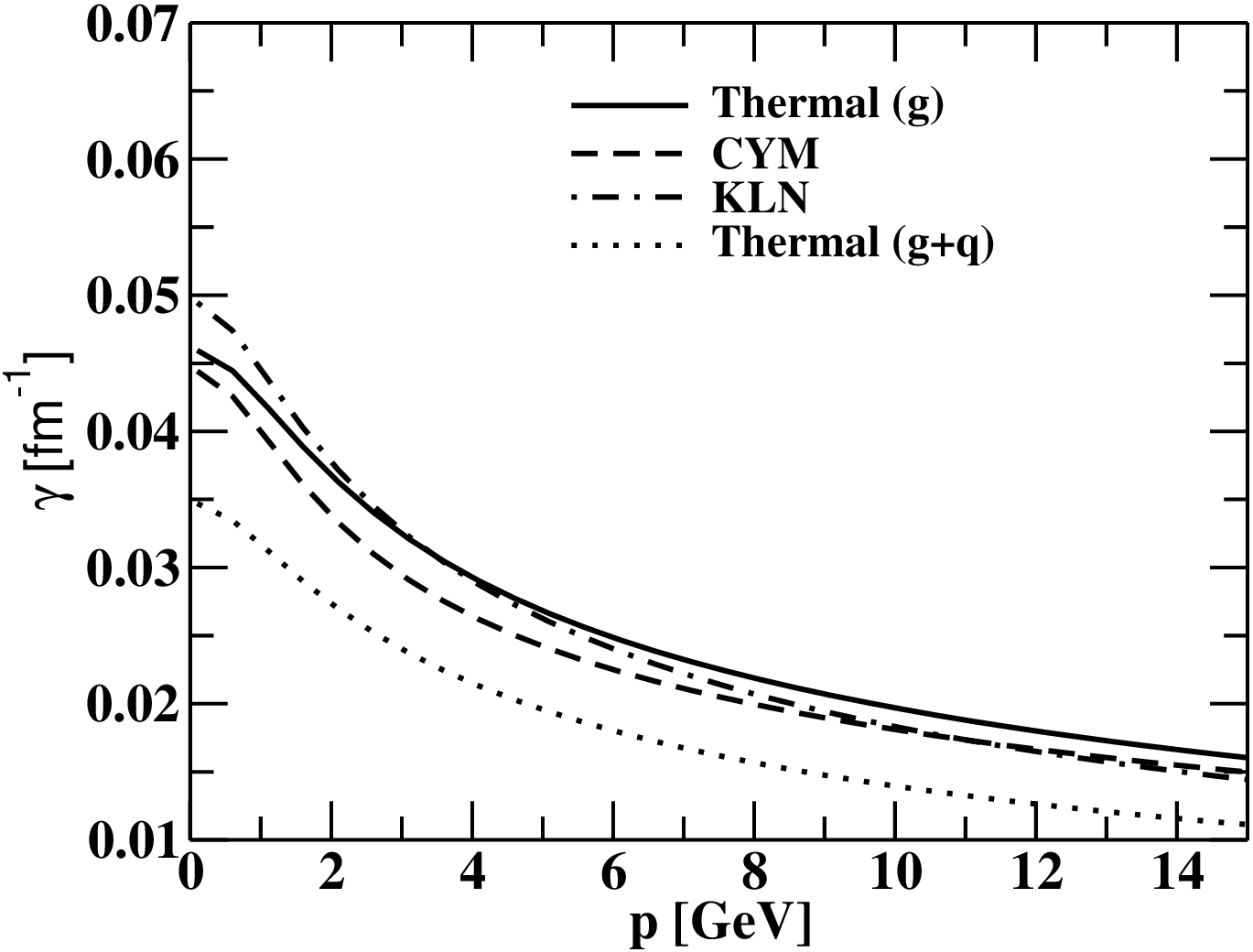}\hspace{2pc}
\includegraphics[width=17pc,clip=true]{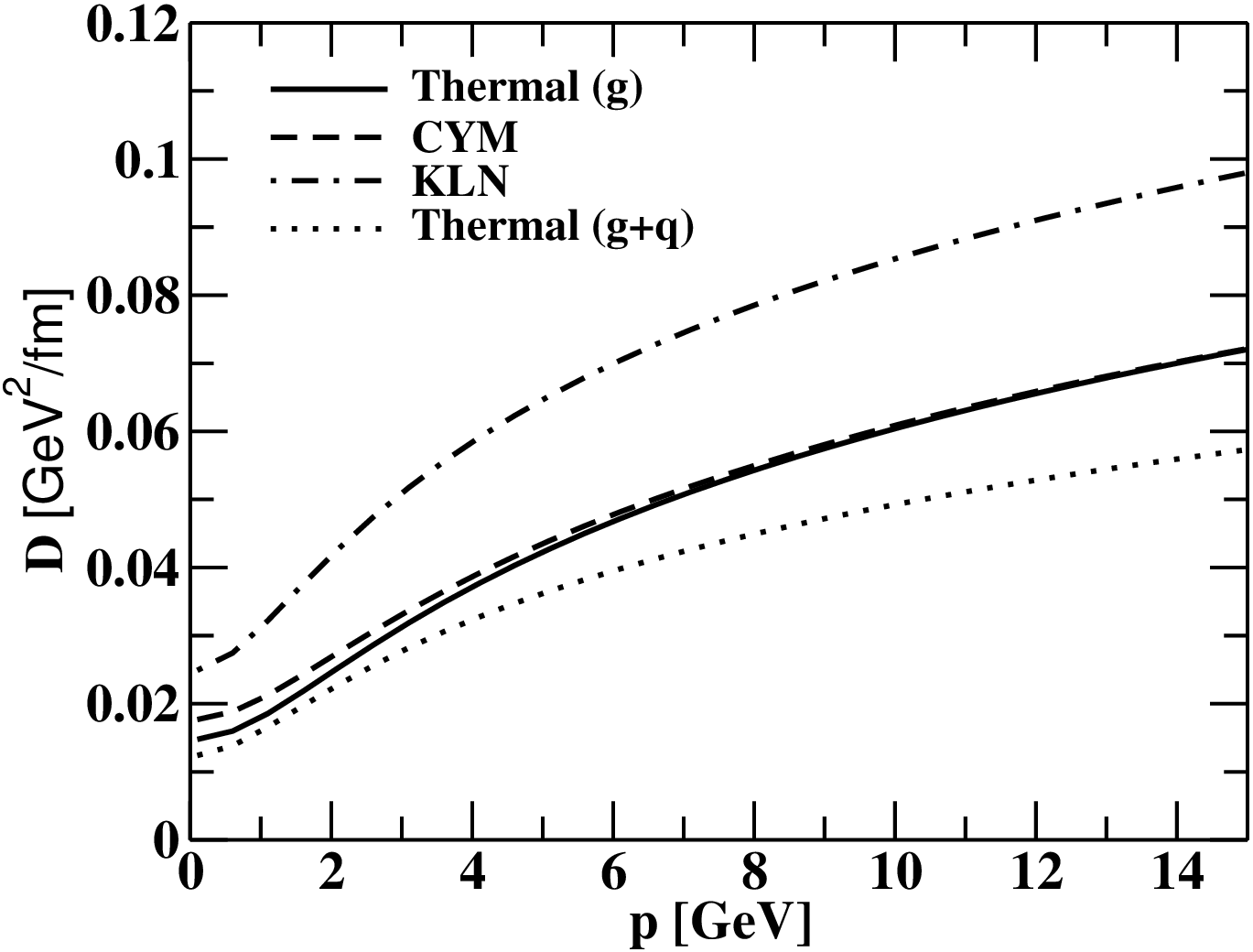}\hspace{2pc}
%\begin{minipage}[b]{14pc}described
\caption{Drag coefficient (left panel) and diffusion coefficient (right panel)
as a function of momentum for charm quark at RHIC energy. 
Debye mass is computed self consistently using Eq.~\ref{mdd}.  
The results corresponding to thermal gluons (kinetic equilibrium) and thermal quarks and 
gluons (chemical equilibrium) are evaluated at a temperature 340 MeV. }
%\end{minipage}
\label{fig3}
\end{center}
\end{figure}
%%%%%%%%%%%%%%%%%%%%%%%%%%%%%%%%%%%%%%%%%%%%%%%%%%%%%%%

Replacing the thermal distribution of  gluons in Eq.~(\ref{mdd}) by 
the KLN and CYM  gluon distributions and setting $f_q=0$ for a gluonic system we estimate $m_D$ and use
this to calculate $\gamma$  and $D$.
The momentum variation of  $\gamma$  and $D$ are displayed in Fig.~\ref{fig3} for charm quarks. 
It is interesting to note that in this case 
$\gamma$ is almost unaffected by the bulk distributions, {\it i.e.}
of KLN, CYM and thermal gluons. We find some difference between the aforementioned bulks and the 
chemically equilibrated QGP, which is caused by the less number of gluons 
in the latter case as discussed earlier.
 
%%%%%%%%%%%%%%%%%%%%%%%%%%%%%%%%%%%%%%%%%%%%%%%%%%%%%%%
\begin{figure}[t!]
\begin{center}
\includegraphics[width=17pc,clip=true]{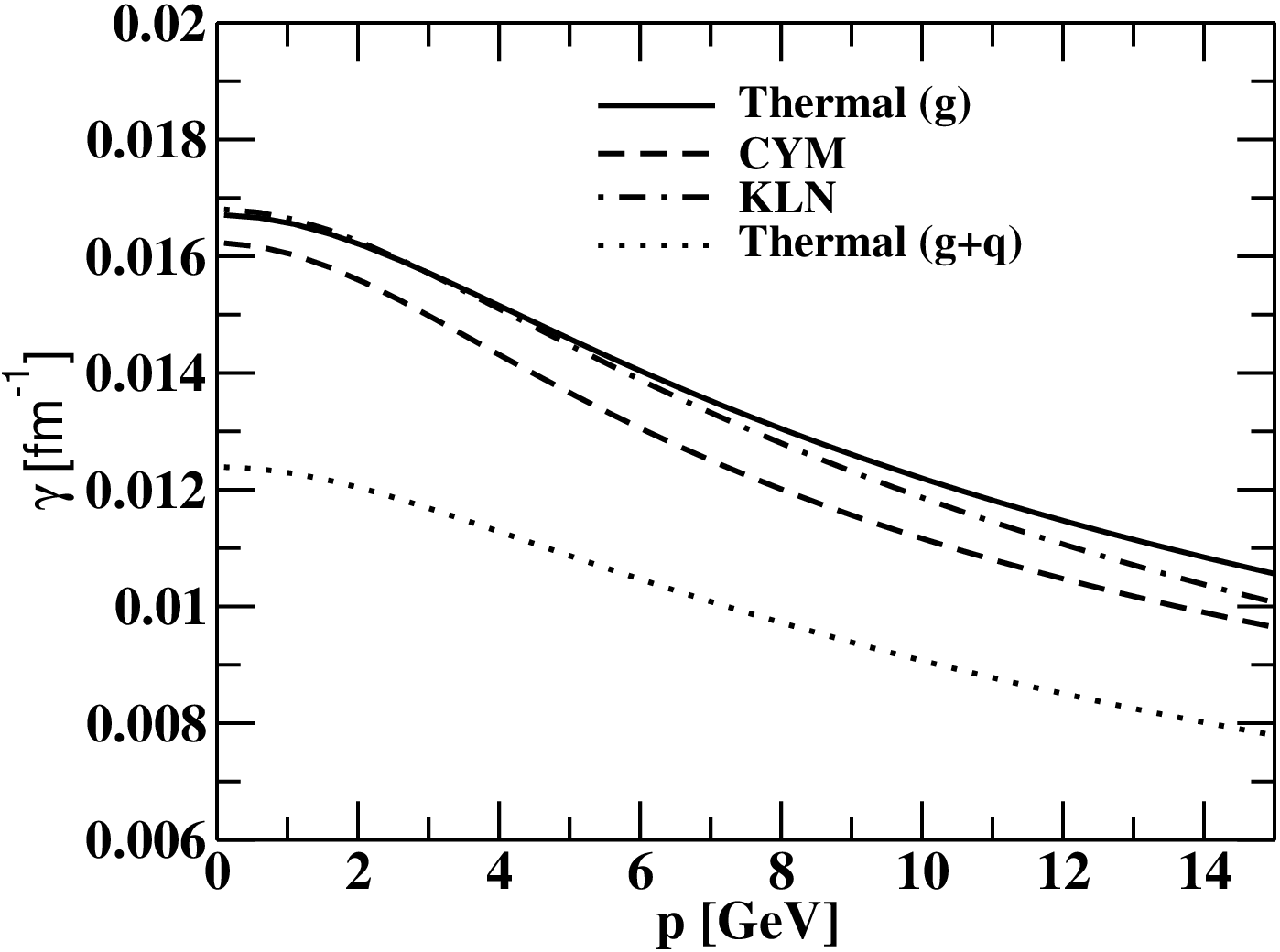}\hspace{2pc}
\includegraphics[width=17pc,clip=true]{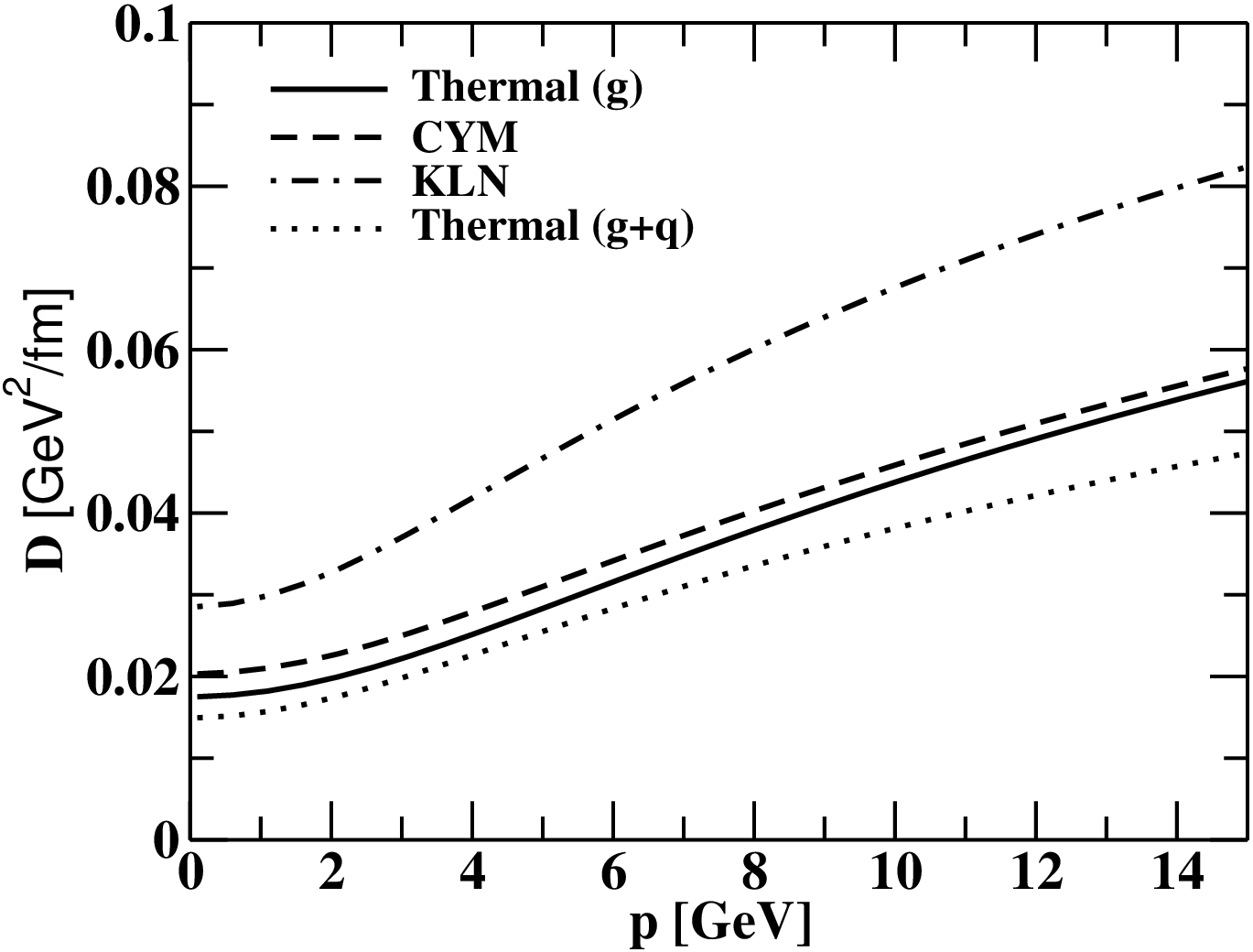}\hspace{2pc}
%\begin{minipage}[b]{14pc}described
\caption{Drag coefficient (left panel) and diffusion coefficient (right panel)
as a function of momentum for bottom quark at RHIC energy. 
Debye mass is computed selfconsistently using Eq.~\ref{mdd}.  
The results corresponding to thermal gluons (kinetic equilibrium) and thermal quarks and 
gluons (chemical equilibrium) are evaluated at a temperature 340 MeV. }
%\end{minipage}
\label{fig5}
\end{center}
\end{figure}
%%%%%%%%%%%%%%%%%%%%%%%%%%%%%%%%%%%%%%%%%%%%%%%%%%%%%%%
 
The drag (left panel) and diffusion (right panel)
coefficients for $b$ quarks in the pre-equilibrium phase
have been displayed in Fig.~\ref{fig5}.  
We notice that these coefficients are smaller for $b$ than $c$.
However, the qualitative variation of $b$ diffusion coefficient 
with momentum is similar to $c$.
 
%%%%%%%%%%%%%%%%%%%%%%%%%%%%%%%%%%%%%%%%%%%%%%%%%%%%%%%
\begin{figure}[t!]
\begin{center}
\includegraphics[width=17pc,clip=true]{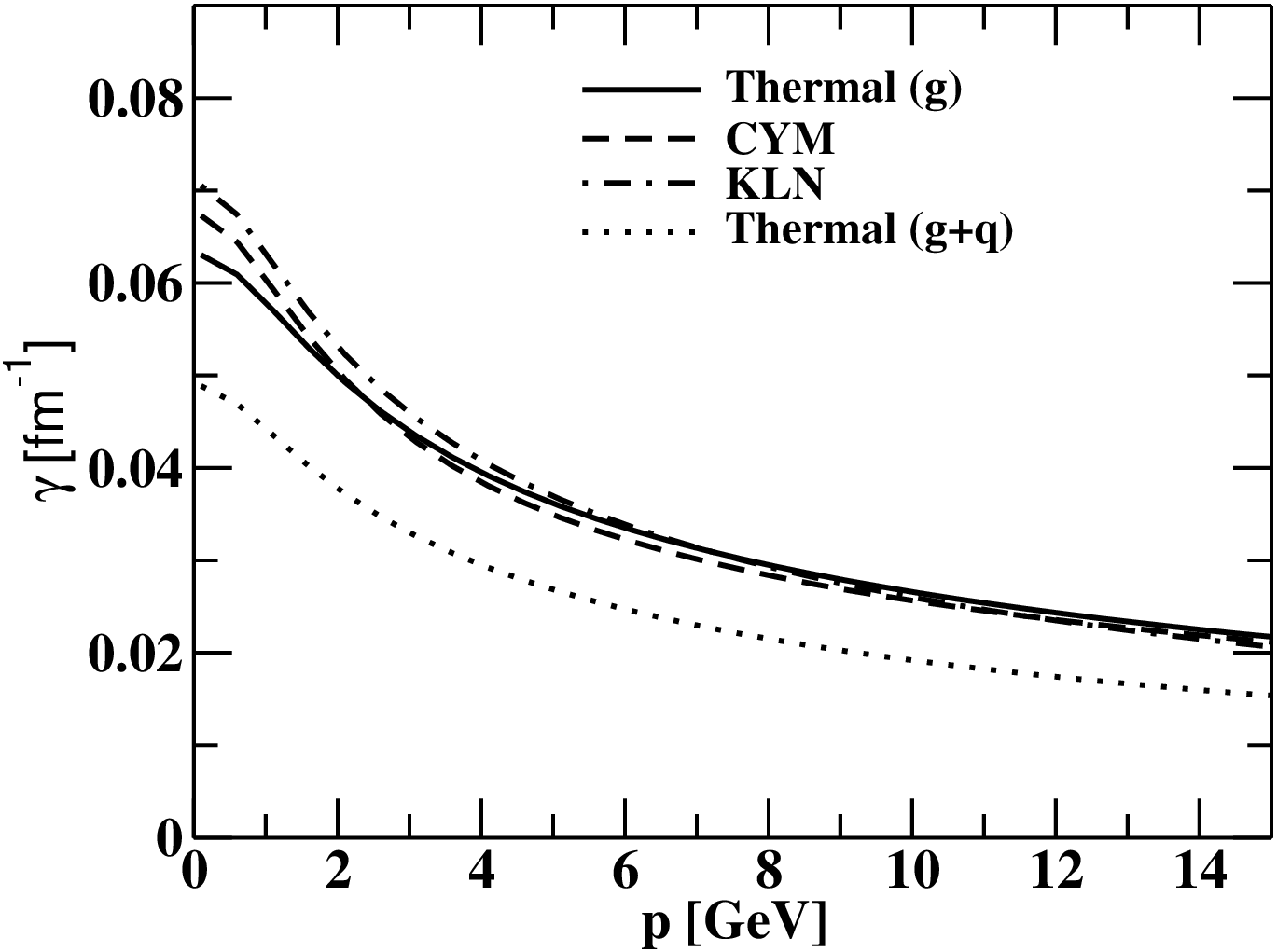}\hspace{2pc}
\includegraphics[width=17pc,clip=true]{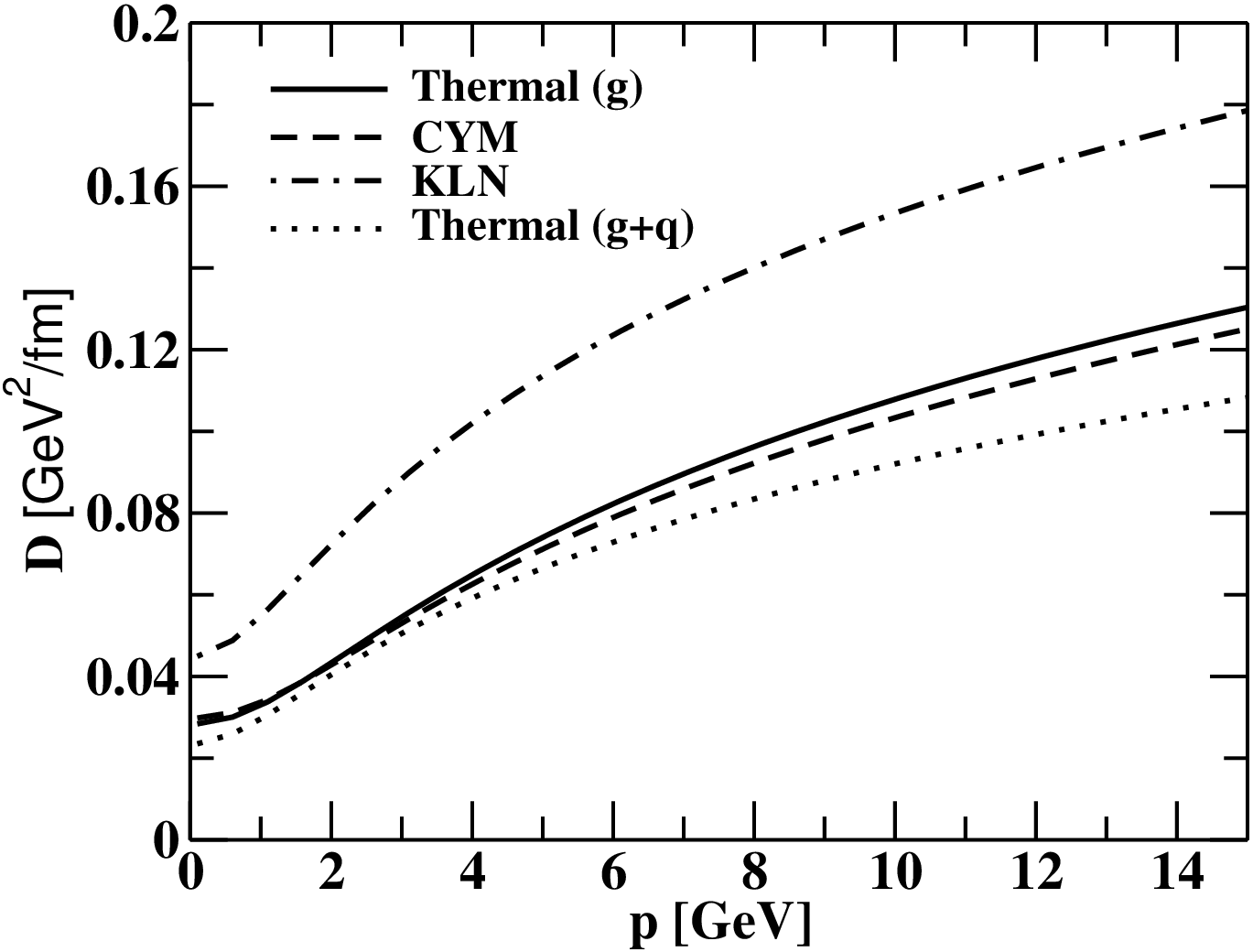}\hspace{2pc}
%\begin{minipage}[b]{14pc}described
\caption{Drag coefficient (left panel) and diffusion coefficient (right panel)
as a function of momentum for charm quark at LHC energy. 
Debye mass is computed selfconsistently using Eq.~\ref{mdd}.  
The results corresponding to thermal gluons (kinetic equilibrium) and thermal quarks and 
gluons (chemical equilibrium) are evaluated at a temperature 510 MeV. }
%\end{minipage}
\label{fig7}
\end{center}
\end{figure}
%%%%%%%%%%%%%%%%%%%%%%%%%%%%%%%%%%%%%%%%%%%%%%%%%%%%%
 
%%%%%%%%%%%%%%%%%%%%%%%%%%%%%%%%%%%%%%%%%%%%%%%%%%%%%%%
\begin{figure}[t!]
\begin{center}
\includegraphics[width=17pc,clip=true]{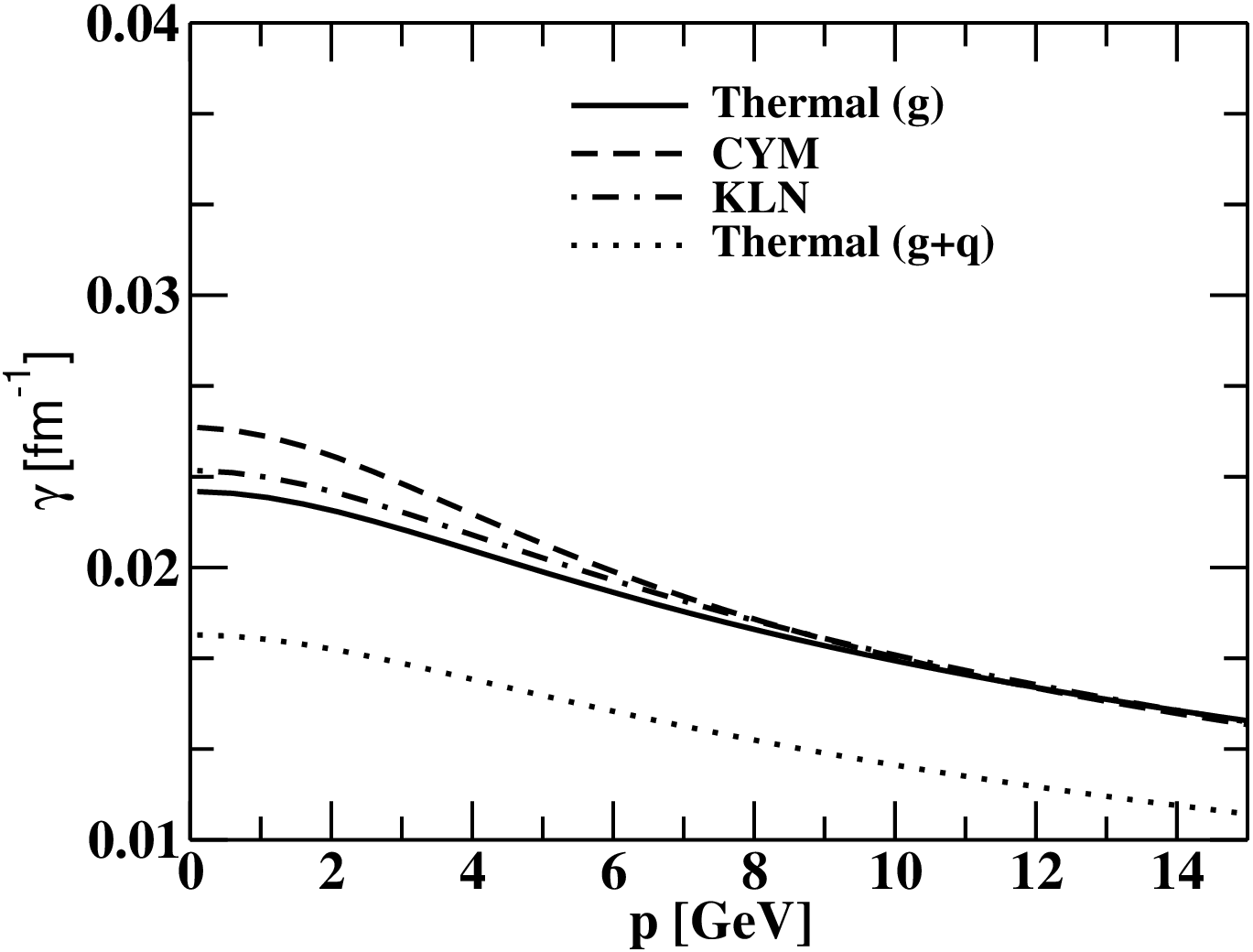}\hspace{2pc}
\includegraphics[width=17pc,clip=true]{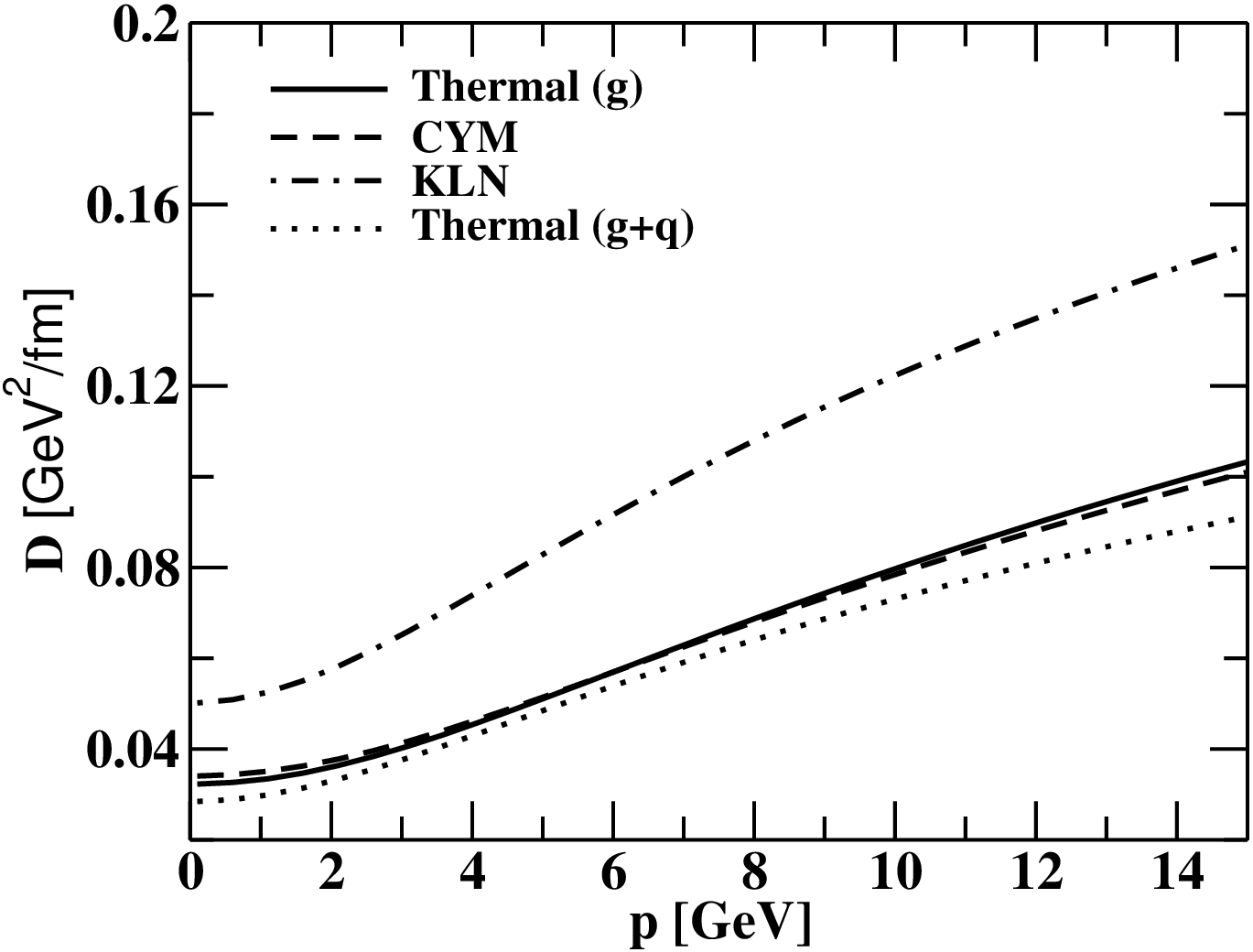}\hspace{2pc}
%\begin{minipage}[b]{14pc}described
\caption{Drag coefficient (left panel) and diffusion coefficient (right panel)
as a function of momentum for bottom quark at LHC energy. 
Debye mass is computed selfconsistently using Eq.~\ref{mdd}.  
The results corresponding to thermal gluons (kinetic equilibrium) and thermal quarks and 
gluons (chemical equilibrium) are evaluated at a temperature 510 MeV. }
%\end{minipage}
\label{fig9}
\end{center}
\end{figure}
%%%%%%%%%%%%%%%%%%%%%%%%%%%%%%%%%%%%%%%%%%%%%%%%%%%%%
 
In the left panel of Fig.~\ref{fig7},  the momentum variation of the $\gamma$ of the $c$ quark                 
in the pre-equilibrium phase is depicted for LHC collision conditions. 
This result is compared with the one computed for a QGP  at temperature T=0.51 GeV. 
%The temperature dependence of the 
%strong coupling $\alpha_s$ is taken from Ref.~\cite{zantow}.
%corresponding to temperature T=0.51 GeV. 
Here same values of the coupling for both the 
equilibrium and pre-equilibrium system have been used. 
The value of $m_D$ is taken from Eq.~(\ref{mdd}) for all the cases. We 
find that the magnitude of the drag 
coefficient in pre-equilibrium phase is similar to that obtained for the kinetic equilibrium case,
however, for a QGP the value of drag is smaller because of the less number of gluons as mentioned earlier. 
In the right panel of Fig.~\ref{fig7} we show the variation of $D$ with $p$ for the $c$ quarks at LHC energy. 
Again the variation of the diffusion coefficient at LHC energy is qualitatively similar to 
RHIC. Similar to RHIC the diffusion at LHC conditions is larger for the KLN distribution.

Similarly the $\gamma$ and $D$ for $b$ quarks 
are plotted in  Fig.~\ref{fig9} for LHC collision conditions.
Qualitatively the  variation of these coefficients with $p$
is similar to  RHIC. Although, the quantitative values 
at LHC collision conditions are larger than RHIC as expected owing 
to the larger density and temperature.. 
 
\section{Summary and discussions}
In this work we have studied the momentum variation of the drag and diffusion coefficients of HQs
in the pre-equilibrium era of  heavy ion collisions at RHIC and LHC energies.
The momentum distribution of the pre-equilibrated gluons have 
been taken from CYM and KLN formalisms. 
This study is motivated by the fact that the effect 
of the pre-equilibrium phase on the HQs suppression and elliptic flow 
might be relevant for low-energy nuclear collisions. 
For example, the simulations of Au+Au collision at RHIC energy
show that the equilibration is achieved approximately within a time scale of $1$ fm/c,
while the lifetime of the QGP phase turns out to be about $5$ fm/c \cite{Ruggieri:2013bda,Ruggieri:2013ova};
hence the system spends about $20\%$ of QGP life-time in the pre-equilibrium phase.
In the case of Pb-Pb collisions at LHC energy the equilibration time
is shorter and lifetime of the QGP phase is larger, hence in this case we  expect
a smaller effect vis-a-vis pre-equilibrium phase. 
%Hence the pre-equilibrium evolution might affect observable like $R_{AA}$. 

We have compared the  magnitudes of the transport coefficients computed 
for equilibrated and pre-equilibrated system, keeping the number 
of particles same in the two cases. We have found that 
the magnitude of the transport coefficients in the pre-equilibrium phase is comparable to
the values obtained with a thermalized gluonic system. 
Moreover, we have also found that the transport coefficients in the pre-equilibrium era are  larger
than the ones obtained for a chemically equilibrated QGP system. This is due to the 
fact that for a fixed number of particles the number of gluons are less 
in the equilibrated QGP than the pre-equilibrated gluonic system and the $HQ+q$ cross section 
is smaller than the $HQ+g$ cross section. 
   
The results obtained in this work may have significant impact on the experimental
observables, for example on heavy mesons $R_{AA}$ and elliptic flow, 
as well as on the suppression of single electron spectra originating from the decays 
of heavy mesons and their elliptic flow. We will address these aspects in future works.

\vspace{2mm}
\section*{Acknowledgments}
We acknowledge B. Schenke and R. Venugopalan for kindly sending us the data
for the CYM spectrum. S.K.D, M.R and V.G acknowledge the support by the ERC StG under the QGPDyn
Grant n. 259684.

%%%%%%%%%%%%%%%%%%%%%%%%%%%%%%%%%%%%%%%%%%%%%%5

\section{References}
%%%%%%%%%%%%%%%%%%%%%%%%%%%%%%%%%%%%%%%%%%%%%%5
  
\end{document}